\documentclass[a4paper,11pt]{article}
\usepackage{pos}
\usepackage{caption}
\usepackage{subcaption}
\usepackage{aas_macros}

\title{Variability studies of active galactic nuclei from the long-term monitoring program with the Cherenkov Telescope Array}
 \ShortTitle{Variability studies of AGN from the long-term monitoring program with CTA}
 
\author*[a]{G.~Grolleron}
\author[b]{J.~Becerra ~Gonzalez}
\author[c]{J.~Biteau}
\author[d]{M.~Cerruti}
\author[e]{R.~Grau}
\author[c]{L.~Gréaux}
\author[f]{T.~Hovatta}
\author[a]{J.-P.~Lenain}
\author[f]{E.~Lindfors}
\author[g]{W.~Max-Moerbeck}
\author[h]{D.~Miceli}
\author[e]{A.~Moralejo}
\author[f]{K.~Nilsson}
\author[k]{E.~Pueschel}
\author[b]{A.~Sarkar}
\author[i,j]{S.~Suutarinen}

\affiliation[a]{Sorbonne Université, CNRS/IN2P3, Laboratoire de Physique Nucléaire et de Hautes Energies, LPNHE, 4 place Jussieu, 75005 Paris, France}

\affiliation[b]{Instituto de Astrofísica de Canarias and Departamento de Astrofísica, Universidad de La Laguna,
La Laguna, Tenerife, Spain}

\affiliation[c]{Laboratoire de Physique des 2 infinis, Irene Joliot-Curie,IN2P3/CNRS, Université Paris-Saclay, Université de Paris,
15 rue Georges Clemenceau, 91406 Orsay, Cedex, France}

\affiliation[d]{Université de Paris, CNRS, Astroparticule et Cosmologie,
10 rue Alice Domon et Léonie Duquet, 75013 Paris Cedex 13, France}

\affiliation[e]{Institut de Fisica d'Altes Energies (IFAE), The Barcelona Institute of Science and Technology,
Campus UAB, 08193 Bellaterra (Barcelona), Spain}

\affiliation[f]{ Finnish Centre for Astronomy with ESO (FINCA), University of Turku,
Vesilinnantie 5, 20014 University of Turku, Finland}

\affiliation[g]{ Departamento de Astronomía, Universidad de Chile,
Camino El Observatorio 1515, Las Condes, Santiago, Chile}

\affiliation[h]{INFN Sezione di Padova and Università degli Studi di Padova,
Via Marzolo 8, 35131 Padova, Italy}

\affiliation[i]{Aalto University Metsähovi Radio Observatory,
Metsähovintie 114, 02540 Kylmälä, Finland}

\affiliation[j]{Aalto University Department of Electronics and Nanoengineering,
PO Box 15500, 00076 Aalto, Finland}

\affiliation[k]{Deutsches Elektronen-Synchrotron, Platanenallee 6,
15738 Zeuthen, Germany}

\onbehalf{for the Cherenkov Telescope Array Consortium} 

\emailAdd{ggroller@lpnhe.in2p3.fr}

\abstract{Blazars are active galactic nuclei (AGN) with a relativistic jet oriented toward the observer. This jet is composed of accelerated particles which can display emission over the entire electromagnetic spectrum. Spectral variability has been observed on short- and long-time scales in AGN, with a power spectral density (PSD) that can show a break at frequencies below the well-known red-noise process. This break frequency in the PSD has been observed in X-rays to scale with the accretion regime and the mass of the central black hole. It is expected that a break could also be seen in the very-high-energy gamma rays, but constraining the shape of the PSD in these wavelengths has not been possible with the current instruments. The Cherenkov Telescope Array (CTA) will be more sensitive by a factor of five to ten depending on energy than the current generation of imaging atmospheric Cherenkov telescopes, therefore it will be possible with CTA to reconstruct the PSD with a high accuracy, bringing new information about AGN variability. In this work, we focus on the AGN long-term monitoring program planned with CTA. The program is proposed to begin with early-start observing campaigns with CTA precursors. This would allow us to probe longer time scales on the AGN PSD.}

\ConferenceLogo{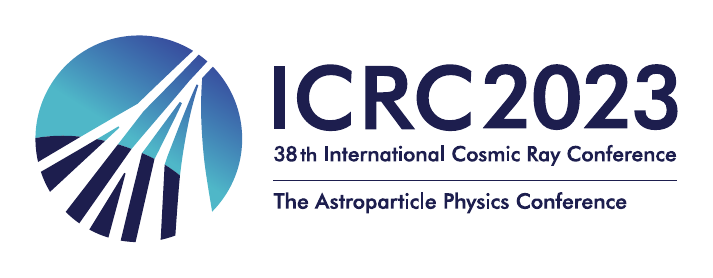}

\FullConference{%
38th International Cosmic Ray Conference (ICRC2023)\\
  26 July - 3 August, 2023\\
  Nagoya, Japan}

\begin{document}
\maketitle

\section[Introduction]{Introduction}
The Cherenkov Telescope Array (CTA) will be the next generation of ground-based imaging atmospheric Cherenkov telescopes (IACT). It will improve the gamma-ray sensitivity above 20 GeV by a factor of five to ten depending on energy compared to current IACT. Therefore, CTA will bring a new outlook on the Universe in particular on its non-thermal emission.

Active Galactic Nuclei (AGN) are astrophysical sources that radiate over the entire electromagnetic spectrum. This emission comes from both thermal and non-thermal processes. One of the most important source of non-thermal emission comes from the relativistic jets, which are located near the central black hole and which can accelerate particles to the relativistic regime. In some AGN, the jet is oriented toward the observer. These objects, called blazars, show temporal variability in both their flux and their spectral properties. AGN variability can be observed on different time scales : AGN flares are responsible to the shortest scale (from minutes to hours) when the long-term variability can be observed on the year timescale \citep{2019NewAR..8701541H}. The short scale AGN behavior can be described by leptonic or hadronic emission, as discussed in a study of AGN flares presented in these proceedings \citep{2023Cerruti}. Measurement of the AGN long-term behavior makes it possible to reconstruct the power spectral density (PSD) that might show a break from pink noise at low frequencies to red noise at high frequencies \citep{2019Galax...7...28R}. This break frequency in the PSD has been observed in X-rays \citep{1999ApJ...514..682E} to scale with the ratio of the mass of the central black hole and the accretion rate \citep{2002MNRAS.332..231U,2006Natur.444..730M}. Since the characterization of the PSD break at very high energies, even the brightest AGN is challenging with current IACT, CTA may be able to reconstruct this break for several of sources.

This proceeding paper focuses on the AGN long-term monitoring program of CTA, which falls within the AGN CTA Key Science Project (KSP) \citep{inbook}. This program consists of the monitoring of 15 selected AGN to be observed over 10 years at least once a week during their visibility period. The goal of this monitoring is to reconstruct in an unbiased way the flux distribution of AGN and the fraction of the time spent above a given flux level (the duty cycle of jetted AGN). Ultimately we also want to accurately estimate the PSD of these sources. This proceeding paper is a follow up of  \citep{2023arXiv230414208C}, where we started to study the prospects of CTA through simulating AGN flares. Here the focus is on a long-term study, using simulated observations of selected AGN over 10 years. From these simulations we evaluate our ability to reconstruct the flux distribution and PSDs using CTA observations. Here we start investigating which sources are the best candidates for high-accuracy PSD reconstruction. In the following sections, we present the procedure to carry this study out. First, the injected AGN temporal and spectral models are generated, then the CTA observations are simulated, and finally a spectral model is fitted to the simulated data to reconstruct the light curve and the PSD.

\section[Creation of AGN models from time series generation]{Creation of AGN models from time series generation}
AGN light curves can be created with a time series generator \citep{2013MNRAS.433..907E}. In X-rays the observed PSD of AGN can follow a pink- to red-noise PSD with a log-normal distribution of the flux \citep{2005MNRAS.359..345U}. At this stage only a pink noise without break is simulated. We use such a model to study the capabilities of CTA to characterize PSDs. The normalization of the PSD depends on the fractional variability $F_{\mathrm{var}}$ parameter which is defined in \citep{2003MNRAS.345.1271V}, the normalization is done through the equation as follows :
\begin{equation}
    F_{\mathrm{var}}= \int_{\frac{1}{T}}^{\frac{1}{2\delta t}} PSD(\nu)d\nu
\label{eq:Fvar}
\end{equation}
where $T = 10\ \mathrm{years}$ is the monitoring campaign duration and $\delta t = 30\ \mathrm{min}$ is the sampling time used fro the generation of the time series. The cadence and the visibility presented in section \ref{Simulation of CTA observations and light curve reconstruction} are considered thereafter. 
For this work, we use $F_{\mathrm{var}} = 2$ to generate a log-normal time series of median equal to one, $\epsilon$, following the procedure described in \citep{2013MNRAS.433..907E}. This value has been set to produce a flux distribution in agreement with those reconstructed for Mrk 421 and Mrk 501, which are the only sources for which it is possible with current IACT \citep{2023arXiv230400835G}.   

We model the time-dependent evolution of the AGN gamma-ray spectra $\Phi(E,t)$ by assuming as spectral shape a log-parabola with exponential cut-offs, yielding:
\begin{equation}
    \Phi_z(E,t) = \Phi(t) \left(\frac{E}{E_0}\right)^{-\Gamma(t)-\beta ln\frac{E}{E_0}} e^{-\frac{E}{E_\mathrm{cut}}} e^{-\tau_{\gamma \gamma}(E,z)}
\label{eq:spectralmodel}
\end{equation}
where $E_0$ is the reference energy, $\Phi(t)$ is the differential flux at the reference energy, $\Gamma(t)$ the spectral index, $\beta$ the log-parabola curvature and $E_\mathrm{cut}$ the cutoff energy. The last factor in Equation \ref{eq:spectralmodel} describes the absorption of VHE photons in the extragalactic background light (EBL) with the optical depth $\tau(E,z)$ depending on the source's redshift, $z$, taken from the work of \citep{2011MNRAS.410.2556D} and the gamma-ray energy. 
In this model $\beta$ and $E_\mathrm{cut}$ are time-independent. Their value, as well as these of the median flux $\Phi_0$ and the spectral index $\Gamma_0$ at $E_0$, are obtained for the AGN within the CTA KSP by joint modelling of archival spectra from \textit{Fermi}-LAT \citep{2022ApJS..260...53A} and IACT \citep{2023Greaux}. Both the flux $\Phi(t)$ and index $\Gamma(t)$ are time-dependent, following a harder-when-brighter behavior \citep{2010A&A...520A..83H,2014MNRAS.444.1077K} : 
\begin{equation}
\begin{split}
    \Phi(t) = \Phi_0 [\epsilon(t)]^{\chi} \\
    \Gamma(t) = \Gamma_0 - b \chi \ln \epsilon(t) \\
\label{eq:flux_index}
\end{split}
\end{equation}
The spectral variability is defined in such a way to have a constant flux at an energy $E_\mathrm{pivot} = 10\ \mathrm{GeV}$ \citep{biteau:pastel-00822242}. To fulfill this condition, we set $b^{-1} = \ln \frac{E_\mathrm{pivot}}{E_0}$ and $\chi^{-1} = 1+b \ln \frac{E_{\mathrm{cut}}}{E_0}$.
\begin{figure*}
    \centering
    \includegraphics[width=0.35\textwidth]{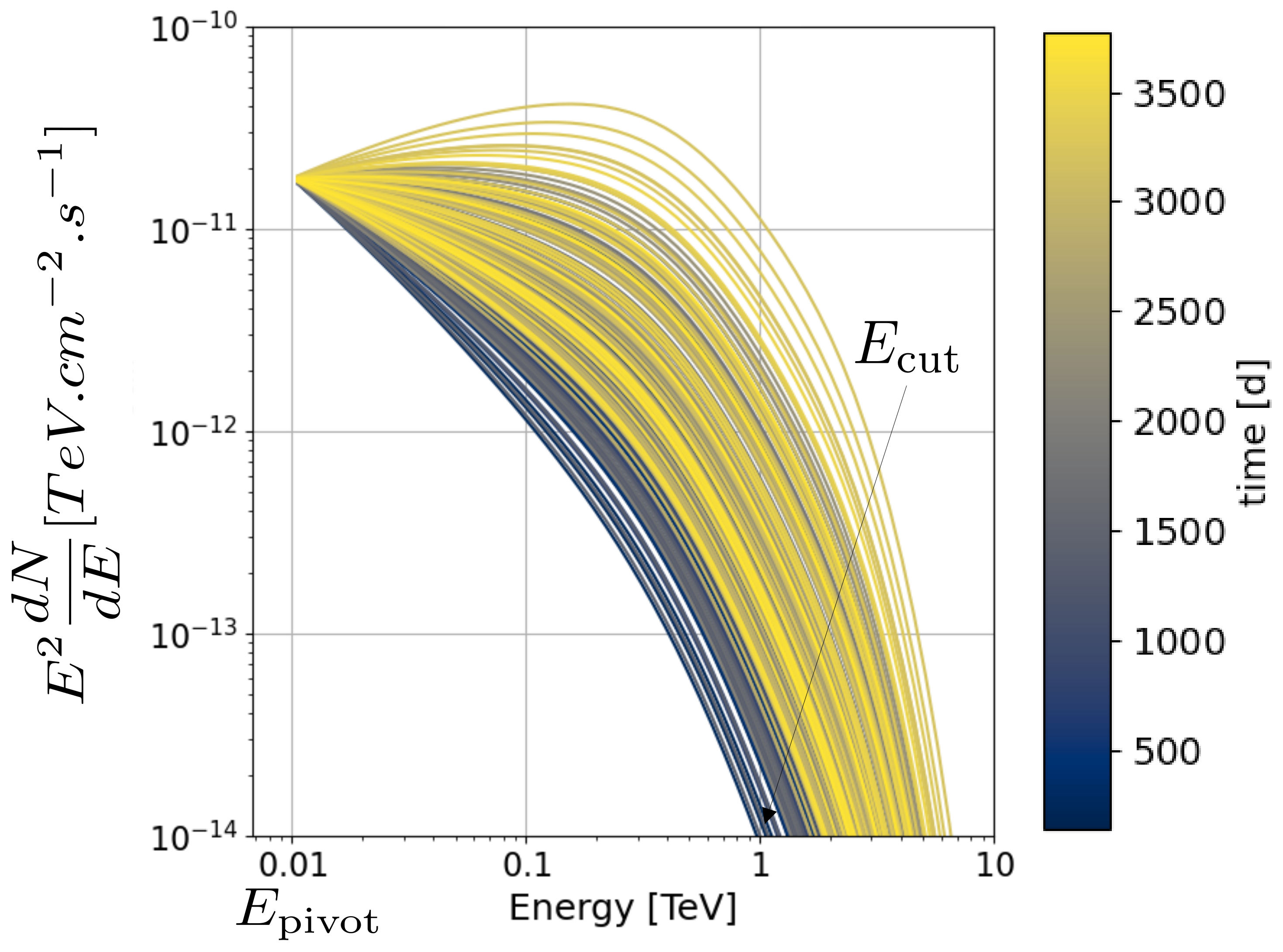}
    \caption{Spectra generated for BL Lac with the time generation algorithm. The color code marks the time evolution for 10 years from blue to red. Obviously, there is no variability at $E_\mathrm{pivot}$ as designed.}
    \label{fig:FigIn}
\end{figure*}
We can see on Figure \ref{fig:FigIn} that the spectra generated for the prototypical blazar BL Lac. The amplitude of the spectral variability, which is described by $F_{\mathrm{var}}$ and is injected at $E_\mathrm{cut}$ through the equation \ref{eq:flux_index}, is clearly visible. We can also see that the spectra are varying around the pivot at $E_\mathrm{pivot}$ where the amplitude of the spectral variability has been set to zero. We will use this model as an input to simulate observations with CTA.

\section[Simulation of CTA observations and light curve reconstruction]{Simulation of CTA observations and light curve reconstruction}
\label{Simulation of CTA observations and light curve reconstruction}
To simulate CTA observations we are using the \textsc{CtaAgnVar}\footnote{\url{https://gitlab.cta-observatory.org/guillaume.grolleron/ctaagnvar}} pipeline which has been designed to simulate and analyse AGN observations with CTA. This package is based on \textsc{Gammapy}, the analysis high-level pipeline for CTA \citep{gammapy17}.  Within the \textsc{CtaAgnVar} pipeline, an observation sequence is created, from an input AGN time-dependent spectral model, taking into account CTA observational constraints and the visibility of the source throughout the night. The tracking of sources allows us to follow the zenith angle and to dynamically select the instrument response functions (IRF) \citep{cherenkov_telescope_array_observatory_2021_5499840}. Thus, it is possible to simulate gamma-like events in a realistic manner and then fit an analytical spectral model. 
Furthermore, the long-term monitoring program is planned to follow the selected AGNs with a cadence of one week with 30 min of integration time. However, as most sources are not circum-polar, they are only visible between 10 to 40 weeks per year depending on their declination.\\
The reconstruction of the light curve is done by fitting a set of spectral models to the simulation. The simplest hypothesis is a power law described by setting $\beta=0$ and $E_{cut}\rightarrow\infty$ in Equation \ref{eq:spectralmodel}. A more complex version is a log-parabola (with $\beta \ne 0$) or a power law with exponential cutoff (a finite $E_{cut}$), and the most complex is the combination of both (log-parabola and power law with exp. cutoff). For each time bin, a more complex model is preferred if the simplest one can be rejected at a 3$\sigma$ level based on a likelihood ratio test. As the log-parabola and the power law with exponential cutoff have the same complexity, the model providing the highest likelihood value is preferred. 
Finally, to quantify the agreement between the data and the model, a goodness-of-fit estimate has been developed based on the likelihood ratio test between the data and the model. The test statistics (TS) derived from the likelihood ratio follows a $\chi 2$ distribution with TS$=0$ describing a perfect fit. We defined therefore a $p_{value}$ to identify possible outliers.

\begin{figure*}
    \centering
    \includegraphics[width = .9\textwidth]{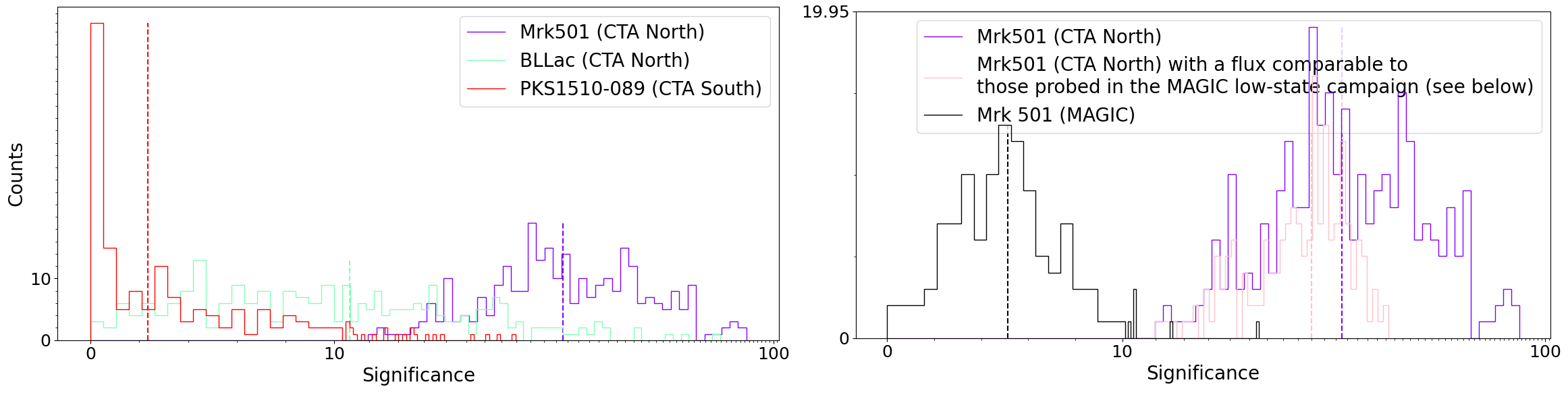}
    \caption{\textit{Left:} Detection significance distribution for 3 promising sources. The vertical dotted line is the median value associated to each source. \textit{Right:} Comparison with the one of Mrk 501 observed by MAGIC in a low-state campaign \citep{Abe_2023}. To allow unbiased comparison between MAGIC data and CTA simulations, the pink distribution is plotted selecting observations with a flux level comparable to the flux distribution observed by MAGIC.}
    \label{fig:Sigma}
\end{figure*}

\begin{figure*}
    \centering
    \includegraphics[width=0.39\linewidth]{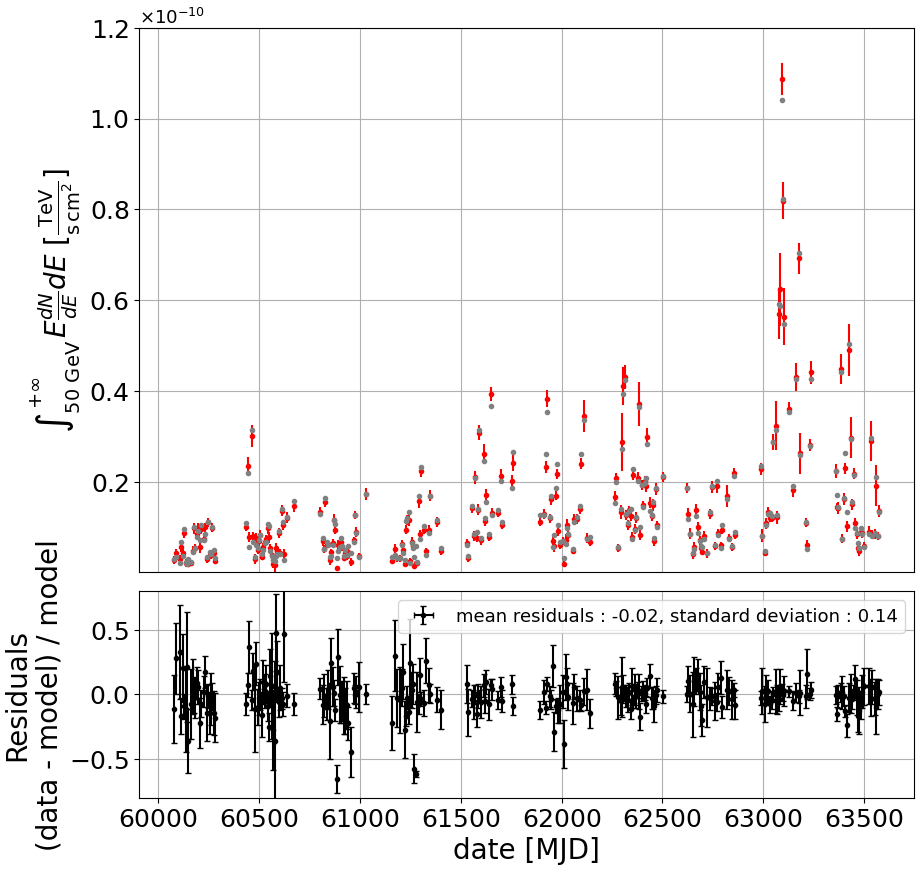}
    \includegraphics[width=0.42\linewidth]{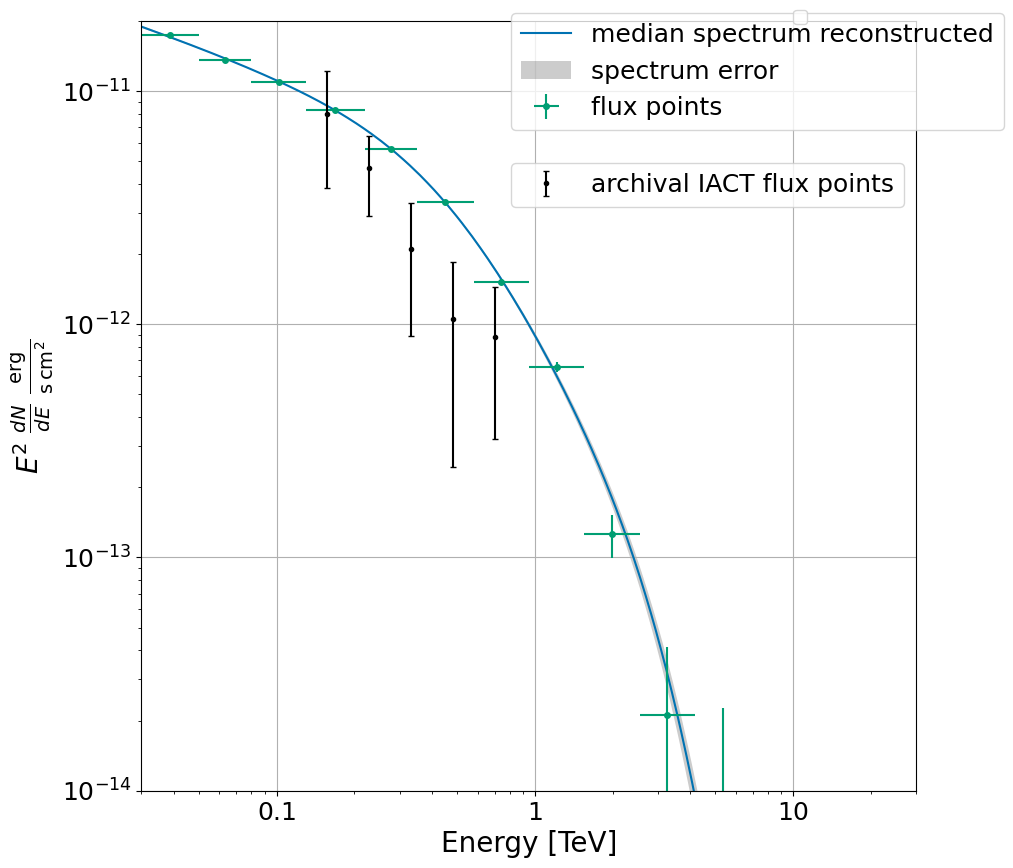}
    \caption{\textit{Left:} Reconstructed light curve and residuals computed between the simulated and reconstructed data for BL Lac for the full simulation of 10 years. Gray points are the injected values and red points are the reconstructed ones. \textit{Right:} Reconstructed median spectrum for BL Lac for the 10 years of data. Spectral parameters are: $\Phi_{0} = 7.46 \pm 0.07 \times 10^{-10}\ \mathrm{TeV^{-1}.s^{-1}.cm^{-2}}$, $\Gamma = 2.37 \pm 0.01$,  $E_\mathrm{cut}^{-1} = 1.08 \pm 0.06\ \mathrm{TeV^{-1}}$ and $\beta$ statistically consistent with 0. Black points are the flux points reconstructed with current IACT obtained from \citep{2023Greaux}}
    \label{fig:FigOut}
\end{figure*}

The simulations have been done for each source listed in the CTA AGN KSP \citep{inbook}. However, here we focus on three bright sources: Mrk 501 (z = 0.033), BL Lac (z = 0.069), and PKS 1510-089 (z = 0.36). The distributions of detection significance in weekly 30-min exposures are shown in Figure  \ref{fig:Sigma}. 
As an example, the reconstructed light curve, for the most promising source with CTA,  BL Lac, above 50 GeV is presented in Figure \ref{fig:FigOut} together with the median spectrum computed over the 10 years of data. The first important result is that there is very good agreement between the injected and the reconstructed light curve with residuals around 1\%.

\section[PSD reconstruction]{PSD reconstruction}
\begin{figure*}
    \centering
    \includegraphics[width=0.31\linewidth]{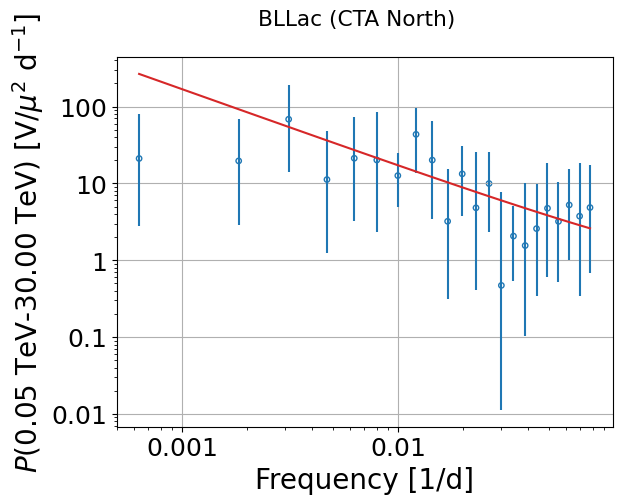}
    \includegraphics[width=0.31\linewidth]{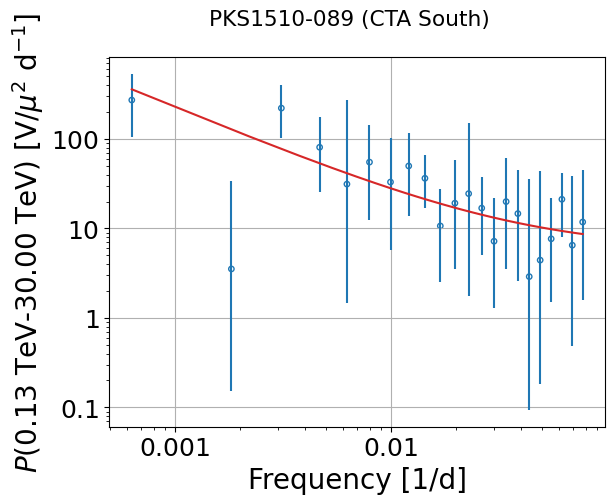}
    \caption{PSDs estimate (blue points) of the simulated data for BL Lac (left) and PKS 1510-089 (right). The red line shows the injection PSD (module floor level at high frequencies) used to simulate the input data.}
    \label{fig:PSD}
\end{figure*}

Having obtained the light curves, we can proceed to estimate, through periodogram, the PSDs for each source following \citep{2013MNRAS.433..907E}. PSDs reconstruction for PKS 1510-089 and BL Lac are presented in Figure \ref{fig:PSD}. The computation with error propagation is performed following the procedure of \citep{biteau:pastel-00822242}. The red line is not a fit of the PSD, but a pink noise spectrum (index of 1) with the Poisson plateau at high frequencies. The plateau is clearly visible in the PSD of PKS 1510-089 (Figure \ref{fig:PSD}, right). The normalization of the red line is done according to Equation \ref{eq:Fvar}. It shows that the source PSDs can be established with the CTA observation scheme, as the red line agrees well with the estimated PSDs.

\section[CTA prospective for long-term AGN studies]{CTA perspective for long-term AGN studies}
In this contribution, we studied CTA's capabilities for a long-term monitoring of AGN. Figure \ref{fig:Sigma} represents the significance distribution for each obervation for several interesting sources in CTA AGN KSP. The detection significance for BL Lac or PKS 1510-089 are comparable to the actual significance distribution of Mrk 501 obtained with current IACT. 

In order to quantify the capabilities of CTA, we have simulated data of various AGN, applied the foreseen observation cadence, and reconstructed the PSDs. While it is possible to reconstruct the input PSD for bright sources, such as BL Lac and PKS 1510-089, the foreseen monitoring schedule (30min per source per week) might need to be optimized for fainter sources. Moreover we can clain that the duty cycle of jetted AGN might be well constrained by CTA. We have just used a simple pink-noise input spectrum here, but will present results employing more complicated spectra in a forthcoming publication. Moreover, this effort will be supported by the observing campaigns with CTA precursors. Nonetheless, the results presented here are encouraging and show that CTA will be able to probe deeply the VHE AGN long-term behavior.

\acknowledgments

This work was conducted in the context of the CTA Consortium. We gratefully acknowledge financial support from the agencies and organizations listed here:
\url{https://www.cta-observatory.org/consortium_acknowledgments/}.

\bibliographystyle{JHEP}
\setlength{\bibsep}{4pt}
\bibliography{skeleton.bbl}

%
%
%

\end{document}